\documentclass[12pt]{article}

\usepackage[round]{natbib} 
\usepackage{rotating, rotfloat}
\usepackage{graphics, graphicx, hyperref, url, amsmath, amsthm, amsfonts, enumerate, epstopdf, subfigure, booktabs, float, color, caption, txfonts}

\usepackage{listings}
\usepackage{color}

\definecolor{dkgreen}{rgb}{0,0.6,0}
\definecolor{gray}{rgb}{0.5,0.5,0.5}
\definecolor{mauve}{rgb}{0.58,0,0.82}

\hypersetup{
    pdfstartview={FitH},    
    colorlinks=true,        
    linkcolor=blue,         
    citecolor=blue,         
    filecolor=blue,         
    urlcolor=blue           
}

     \def\etal{{\itshape et al.}}
\def\beqn{\begin{eqnarray}} \def\eeqn{\end{eqnarray}}

\newcommand{\bec}{\begin{center}}
\newcommand{\enc}{\end{center}}
\newcommand{\bee}{\begin{eqnarray*}}
\newcommand{\ene}{\end{eqnarray*}}
\newcommand{\beq}{\begin{equation}}
\newcommand{\eeq}{\end{equation}}

\newtheorem{algorithm}{Algorithm}

\begin{document}

\title{\textbf{Fast implementation of the Tukey depth}
\thanks{Corresponding author's email: csuliuxh912@gmail.com}}

\author {{Xiaohui Liu$^{a, b}$}\\
         {\em\footnotesize $^a$ School of Statistics, Jiangxi University of Finance and Economics, Nanchang, Jiangxi 330013, China}\\
         {\em\footnotesize $^b$ Research Center of Applied Statistics, Jiangxi University of Finance and Economics, Nanchang, Jiangxi 330013, China}\\
}

\date{}
\maketitle

\begin{center}
\begin{minipage} {12.5cm}{ 
\vspace*{-7mm} \small \textbf{Abstract}.
\small
Tukey depth function is one of the most famous multivariate tools serving robust purposes. It is also very well known for its computability problems in dimensions $p \ge 3$. In this paper, we address this computing issue by presenting two combinatorial algorithms. The first is naive and calculates the Tukey depth of a single point with complexity $O\left(n^{p-1}\log(n)\right)$, while the second further utilizes the quasiconcave of the Tukey depth function and hence is more efficient than the first. Both require very minimal memory and run much faster than the existing ones. All experiments indicate that they compute the exact Tukey depth.
\vspace{1mm}

{\small {\bf\itshape Key words:} Tukey depth; Quasiconcave; Combinatorial property; Fast computation} \vspace{1mm}

{\small {\bf2000 Mathematics Subject Classification Codes:} 62F10; 62F40; 62F35}}
\end{minipage}
\end{center}

\setlength{\baselineskip}{1.25\baselineskip}

\vskip 0.1 in
\section{Introduction}
\paragraph{}
\vskip 0.1 in

To provide a desirable ordering for multivariate data, \cite{Tuk1975} heuristically proposed the useful tool of statistical depth function. With respect to the distribution $P$ of $X$ in $\mathcal{R}^p$ ($p \ge 1$), he defined the Tukey depth of a point $z$ as the minimum probability mass carried by any closed halfspace containing $z$. That is, 
\begin{eqnarray*}
    D(z|P) = \inf_{u \in \mathcal{S}^{p - 1}} P(u^{\top} X \leq u^{\top} z),
\end{eqnarray*}
where $\mathcal{S}^{p - 1} = \{v\in R^{p}: \|v\| = 1\}$. For $n$ $p$-variate observations $\mathcal{X}^n := \{X_{i}\}_{i=1}^n$, its sample version is correspondingly
\begin{eqnarray}
\label{HDSV}
D_{n} (z) := D(z|P_n) = \inf_{u \in \mathcal{S}^{p - 1}} P_n (u^{\top} X \leq u^{\top} z),
\end{eqnarray}
where $P_n$ denotes the empirical distribution of $\mathcal{X}^n$.
\medskip

The Tukey depth has proved very desirable. It satisfies all four properties that define a general notion of statistical depth functions, namely, affine invariance, maximality at center, monotonicity relative to deepest point, and vanishing at infinity \citep{ZS2000}. In practice, it finds many applications in cases such as confidence region constructions \citep{YS1997} and classifications \citep{LCL2012}. Under mild conditions, it even characterizes the underlying distribution \citep{KZ2010}. Latest developments indicate that the Tukey depth has a strong connection with the multiple-output quantile regression methodology \citep{HPS2010, KM2012}.
\medskip

However, its exact computation is challenging. This is mainly because: $P_n (u^{\top} X \leq u^{\top} z)$ is discontinuous and non-convex with respect to $u \in \mathcal{S}^{p-1}$, while $\mathcal{S}^{p-1}$ contains a infinite number of $u$. Hence, it is difficult to find the infimum of $P_n (u^{\top} X \leq u^{\top} z)$ through conventional optimization methods. To be computable, special attention should be paid first to the reduction of the number of $u$. Excellent works in that direction are pioneered by \cite{RR1996a} for \emph{bivariate} data and \cite{RS1998} for 3-dimensional data, respectively, relying on the idea of a circular sequence \citep{Ede1987}. 
\medskip

For \emph{data in spaces of dimension $p > 2$}, \cite{LZ2014a} developed a feasible cone enumeration procedure based on the breadth-first search algorithm. The cones considered by \cite{LZ2014a} satisfy that their vertexes contain all ${n \choose p - 1}$ \emph{critical direction vectors}, which are normal to the hyperplanes passing through $\{z,\, X_{i_1},\, \cdots,\, X_{i_{p-1}}\}$, where $i_1,\, \cdots, i_{p-1}$ are distinct and $i_1,\, \cdots, i_{p-1} \in \{1,\,2,\, \cdots,\, n\}$. Recently, \cite{DM2014} further refined the algorithm of \cite{LZ2014a}. He found that it is possible to calculate the Tukey depth by directly considering these ${n \choose p - 1}$ critical direction vectors. Since his approach is of combinatorial nature, and needs not to take account of any space ordering, his implementation requires much less memory and runs much faster.
\medskip

In this paper, we further improve Mozharovskyi's procedure. We find that it is convenient to extend the definition of the Tukey depth for a single point $z$ into the version for a subspace $\mathbf{V}$. Then relying on this, we propose for dimensions $p \ge 3$ our first exact algorithm which is still of combinatorial nature, but possesses exactly the complexity of $O(n^{p-1} \log n)$, better than that $O(n^p)$ of \cite{DM2014}.
\medskip

Nevertheless, likewise to all algorithms aforementioned, this algorithm still needs to fully address all ${n \choose p - 1}$ critical direction vectors. On the other hand, when computing the Tukey depth, we are in fact searching for the infimum of $P_n (u^{\top} X \leq u^{\top} z)$ with respect to $u$. A great proportion of critical direction vectors may be \emph{redundant} in the sense that some of them have values $P_n (u^{\top} X \leq u^{\top} z)$ larger than $\tau$, which we assume to be an upper bound for the Tukey depth obtained through an approximate method. A natural question that arises now is whether we can eliminate some of them from consideration.
\medskip

The answer is positive. With the extended definition above, we find it is possible to utilize the \emph{quasiconcave}, i.e., all depth regions are convex and nested \citep{Mos2013}, of the Tukey depth function to reduce greatly the number of critical direction vectors involved. An iterative algorithm is constructed to realize this idea. 
This approach is still of combinatorial property because it is strictly limited to consider critical direction vectors. Hence its implementation runs quite efficiently. This algorithm is depth-depending. The smaller the Tukey depth of $z$ is, the less time this algorithm tends to consume.
\medskip

Both algorithms have been implemented in Matlab. The whole code can be obtained from the author through email. Data examples are also provided to illustrate the performance of the proposed algorithms.
\medskip

The rest of this paper is organized as follows. Section \ref{TDSubspace} extends the conventional definition of the Tukey depth for a single point to the version for a subspace. Section \ref{ReCombHD} provides a refined combinatorial algorithm for exactly computing the Tukey depth. Section \ref{ADIA} develops an adaptively iterative procedure. Several data examples are given in Section \ref{Comparisons} to illustrate the performance of the proposed algorithms. Section \ref{Conclude} ends the current paper with a few concluding discussions.

\vskip 0.1 in
\section{Tukey depth for a subspace}
\paragraph{}
\vskip 0.1 in
\label{TDSubspace}

In the literature, it's known that it is difficult to utilize some information, such as quasiconcave, of the Tukey depth function to improve the efficiency of the algorithms constructed directly on \eqref{HDSV}.
To this end, we propose to consider the following extended version of \eqref{HDSV}.
\medskip


Note that $D(z|P_n) = D(0|P_{nz})$ holds for any given $z \in \mathcal{R}^p$ by the affine invariance, in the sequel we suppose that $z = 0$, and pretend the real observations to be $\mathbf{X}^n := \{\mathbf{x}_i\}_{i=1}^n$, where $\mathbf{x}_i = X_{i} - z$, and $P_{nz}$ denotes the empirical distribution of $\mathbf{X}^n$. For convenience, we assume that $\mathbf{X}^n \cup \{0\}$ are in general position, which is common in the literature concerning statistical depth functions; see, e.g., \cite{DG1992} and \cite{MLB2009}. (If the data are not in general position, the subsequent discussions and algorithms need to be modified, e.g., by slightly perturbing the data.)
\medskip

Let $\mathbf{W}^{\bot}$ be the orthogonal complement of the subspace $\mathbf{W}$. Then for a $r$-dimensional subspace $\mathbf{V}_{r}$ of $\mathcal{R}^p$ ($0 \leq r < p$), we define its Tukey depth with respect to $\mathbf{X}^n$ as follows:
\begin{eqnarray}
\label{HDsapce}
    D_{n}(\mathbf{V}_{r}) = \inf_{u \in \mathcal{S}^{p-1} \cap \mathbf{V}_{r}^{\bot}} P_n(u^{\top} X \leq 0).
\end{eqnarray}
When $r = 0$, we assume that $\mathbf{V}_r$ contains only a single point $\{0\}$, and its orthogonal complement subspace is the whole $\mathcal{R}^p$. In this sense, $D_{n}(\mathbf{V}_{r})$ may be referred to as an extension of \eqref{HDSV}.
\medskip

Clearly, for a given $\mathbf{V}_{r}$ ($0 < r < p$), it holds $D_{n}(\mathbf{V}_{r}) \ge D_{n}(0)$. Based on this, it is trivially that
\begin{eqnarray}
\label{HDhyperplane}
    D_{n}(0) = \inf_{\mathbf{V}_{r} \in \mathcal{V}_{r}} D_{n}(\mathbf{V}_{r}),
\end{eqnarray}
where $\mathcal{V}_{r}$ denotes the set containing all $r$-dimensional subspaces. When $r = p - 1$, \eqref{HDhyperplane} deduces to
\begin{eqnarray*}
    D_{n}(0) = \inf_{u \in \mathcal{S}^{p-1}} D_{n}(\mathbf{H}_{u})
\end{eqnarray*}
with $\mathbf{H}_{u} = \{x \in \mathcal{R}^p \ |\ u^\top x = 0\}$ for $u \in \mathcal{S}^{p-1}$.
\medskip


When $\mathbf{X}^n \cup \{0\}$ are in general position, \cite{DM2014} have recently showed that ${n \choose p - 1}$ critical director vectors suffice for computing exactly the Tukey depth; see Algorithm 5.3 and Corollary 5.3 of \cite{DM2014}. This in fact implies that, from the point of view of subspaces, ${n \choose p - 1}$ subspaces $\mathbf{S}_{\mathbf{I}_{p-1}}$ spanned by $p - 1$ points in the sample are sufficient to determine $D_{n}(0)$. That is, he actually obtained
\begin{eqnarray}
\label{DNPp1}
    D_{n}(0) = \min_{\mathbf{I}_{p-1} \in \mathcal{I}_{p-1}} D_{n}(\mathbf{S}_{\mathbf{I}_{p-1}}) - \frac{p - 1}{n},
\end{eqnarray}
where $\mathcal{I}_{p-1}$ is specified in \eqref{indxset}. This result is actually a special case of the following proposition.
\medskip

\textbf{Proposition 1}. \textit{Assume that $\mathbf{X}^n \cup \{0\}$ are in general position. Then for any $r = 1, 2, \cdots, p - 1$, we have that
\begin{eqnarray*}
    D_{n}(0) = \min_{\mathbf{I}_{r} \in \mathcal{I}_{r}} D_{n}(\mathbf{S}_{\mathbf{I}_{r}}) - \frac{r}{n},
\end{eqnarray*}
where}
\begin{eqnarray}
\label{indxset}
    \mathcal{I}_{r} = \{\{i_1,\, i_2,\, \cdots,\, i_{r}\} \ |\ i_1,\, i_2,\, \cdots,\, i_r \text{ distinct, and }i_1,\, i_2,\, \cdots,\, i_r \in \{1,\, 2,\, \cdots,\, n\}\},
\end{eqnarray}
\textit{and $\mathbf{S}_{\mathbf{I}_{r}} := \mathbf{S}_{i_1,\, i_2,\, \cdots,\, i_{r}} = \text{span}(\mathbf{x}_l: l = i_1,\, i_2,\, \cdots,\, i_r)$ denotes the $r$-dimensional subspace of $\mathcal{R}^p$ spanned by $\{\mathbf{x}_{i_1},\, \mathbf{x}_{i_2},\, \cdots,\, \mathbf{x}_{i_r}\}$.}
\medskip

\textbf{Proof}. For a given $\mathbf{I}_{r} := \{i_1,\, i_2,\, \cdots,\, i_{r}\} \in \mathcal{I}_{r}$, 
let $X = X' + X''$, where $X' \in \mathbf{S}_{\mathbf{I}_r}$ and $X'' \in \mathbf{S}_{\mathbf{I}_r}^{\bot}$, and for any $x \in \mathcal{R}^{p}$, let $x^{*} = (x^{\top} \xi_1,\, x^{\top} \xi_2,\,\cdots,\, x^{\top} \xi_{p-r})^\top$, where $\{\xi_1,\, \xi_2,\, \cdots,\, \xi_{p-r}\}$ denotes a standard orthogonal basic of $\mathbf{S}_{\mathbf{I}_r}^\bot$. (Under the assumption of this proposition, the affine dimension of $\mathbf{S}_{\mathbf{I}_r}$ is $r$.) Then the fact, that $u^{\top} X = u^{\top} X'' = (u^*)^{\top} X^*$ holds for any $u \in \mathcal{S}^{p-1} \cap \mathbf{S}_{\mathbf{I}_r}^{\bot}$, implies that
\begin{eqnarray*}
    D_{n}(\mathbf{S}_{\mathbf{I}_r}) = \inf_{u \in \mathcal{S}^{p-1} \cap \mathbf{S}_{\mathbf{I}_r}^{\bot}} P_n(u^{\top} X'' \leq 0) = \inf_{v \in \mathcal{S}^{p - r - 1}} P_n^{*}(v^{\top} X^{*} \leq 0) =: D_{n, {\mathbf{I}_r}}(0), 
\end{eqnarray*}
where $P_n^{*}$ denotes the empirical distribution function of $\{\mathbf{x}^*| \mathbf{x} \in \mathbf{X}^n\}$. That is, one can deduce the computation of $D_{n}(\mathbf{S}_{\mathbf{I}_r})$ into the issue of calculating $D_{n, {\mathbf{I}_r}}(0)$ in the lower-dimensional space.
\medskip

Write $\mathcal{M} = \{1,\, 2,\, \cdots,\, n\} \setminus \{i_1,\, i_2,\, \cdots,\, i_r\}$. Denote $\widetilde{D}_{n, {\mathbf{I}_r}}(0) := \inf_{v \in \mathcal{S}^{p - r - 1}} \widetilde{P}_n^{*}(v^{\top} X^{*} \leq 0)$ with $\widetilde{P}_n^{*}$ being the empirical distribution function of $\{\mathbf{x}_l^*| l \in \mathcal{M}\}$. Note that $\mathbf{x}_{i_1}^* = \mathbf{x}_{i_2}^* = \cdots = \mathbf{x}_{i_r}^* = 0$. Hence,
\begin{eqnarray}
\label{DD1}
    (n - r) \times \widetilde{D}_{n, {\mathbf{I}_r}}(0) + r = n \times D_{n, {\mathbf{I}_r}}(0),
\end{eqnarray}
and, for each $\mathbf{J}_{p - r - 1} \in \mathcal{J}_{p - r - 1}$,
\begin{eqnarray}
\label{DD2}
    (n - r) \times \widetilde{D}_{n, {\mathbf{I}_r}}(\mathbf{S}_{\mathbf{J}_{p - r - 1}}^{*}) + r = n \times D_{n, {\mathbf{I}_r}}(\mathbf{S}_{\mathbf{J}_{p - r - 1}}^{*}).
\end{eqnarray}
Here $\mathcal{J}_{p-r-1} = \{\{j_1,\, j_2,\, \cdots,\, j_{p-r-1}\} \ |\ j_1,\, j_2,\, \cdots,\, j_{p-r-1}$ distinct, and $j_1,\, j_2,\, \cdots,\, j_{p-r-1} \in \mathcal{M}\}$, and $\mathbf{S}_{\mathbf{J}_{p - r - 1}}^{*} = \text{span}(\mathbf{x}_l^{*}: l = j_1,\, j_2,\, \cdots,\, j_{p-r-1})$ denotes the subspace spanned by $\{\mathbf{x}_{j_1}^{*}$, $\mathbf{x}_{j_2}^{*},\, \cdots,\, \mathbf{x}_{j_{p-r-1}}^{*}\}$.
\medskip

Next, for $l \in \mathcal{M}$, let $\Pi_l = \{v \in \mathcal{R}^{p-r} | v^\top \mathbf{x}_{l}^{*} = 0\}$. Clearly, $\Pi_{l_1} \neq \Pi_{l_2}$ if $l_1 \neq l_2$ for $l_1$, $l_2\in \mathcal{M}$ under the in-general-position assumption. (Otherwise, there exists a ($p-1$)-dimensional affine space containing at least $p + 1$ observations. This contradicts with the assumption.) This implies that, for any $\mathbf{J}_{p - r - 1} \in \mathcal{J}_{p - r - 1}$, the affine dimension of $\mathbf{S}_{\mathbf{J}_{p - r - 1}}^{*}$ is always equal to $(p-r)-1$. Hence, Steps 3a and 3e of Algorithm 5.3 in \cite{DM2014} are never true, and a similar proof to that of Theorem 5.2 in \cite{DM2014} guarantees that
\begin{eqnarray*}
    \widetilde{D}_{n, {\mathbf{I}_r}}(0) = \min_{\mathbf{J}_{p - r - 1} \in \mathcal{J}_{p - r - 1}} \widetilde{D}_{n, {\mathbf{I}_r}}(\mathbf{S}_{\mathbf{J}_{p - r - 1}}^{*}) - \frac{(p - r) - 1}{n - r}.
\end{eqnarray*}
This, together with \eqref{DD1} and \eqref{DD2}, leads to
\begin{eqnarray*}
    \min_{\mathbf{I}_r \in \mathcal{I}_r} D_{n}(\mathbf{S}_{\mathbf{I}_r}) = \min_{\mathbf{I}_r \in \mathcal{I}_r} \left\{\min_{\mathbf{J}_{p - r - 1} \in \mathcal{J}_{p - r - 1}} D_{n, {\mathbf{I}_r}}(\mathbf{S}_{\mathbf{J}_{p - r - 1}}^{*}) - \frac{p - r - 1}{n}\right\}
    = D_{n}(0) + \frac{r}{n}.
\end{eqnarray*}
Then this proposition follows immediately.
\medskip

Proposition 1 coincides with the result obtained by \cite{LZ2014a}. That is, for $u_0 \in \mathcal{S}^{p-1}$ such that
\begin{eqnarray*}
    D_{n}(0) = \inf_{u\in \mathcal{S}^{p-1}} P_n(u^{\top} X \leq 0) = P_n(u_0^{\top} X \leq 0),
\end{eqnarray*}
the hyperplane $\{x\in \mathcal{R}^p\ |\ u_0^{\top} x = 0\}$ contains no observation of $\mathbf{X}^n$ when $\mathbf{X}^n \cup \{0\}$ are in general position.
\medskip

Furthermore, it is worth mentioning that special attention should be paid to the adjusted term $-\frac{r}{n}$ (or $r$) when constructing algorithms based on the critical direction vectors. Omitting such a term would lead the Tukey depth to be \emph{overestimated} in the sense that the computed depth value would be strictly greater than the true one no matter how many random direction vector are utilized. Examples can be found in the literature such as \cite{RS1998}; see the \emph{third} approximation algorithm in Page 196. Over there, they investigated a data set that consists of $86$ observations of dimension $p = 8$. The true depth value of $\theta_1$ with respect to this data set is $\leq 16/86$, while that of the second point $\theta_2$ is $0$. From Table 1 of this paper, we can see that the approximate depth values of both points computed through the third approximation algorithm are much greater than $16/86$ and 0, respectively.  (Each value in Table 1 dividing by $n=86$ is correspondingly equal to the approximate depth value.) However, if further subtracting the value $(p-1)/86 = 7/86$, this method would appear to perform much better than what has been reported in Example, as well as Table 1, in Page 196 of \cite{RS1998}.

\vskip 0.1 in
\section{A refined combinatorial algorithm}
\paragraph{}
\vskip 0.1 in
\label{ReCombHD}

Since the computation of the Tukey depth is trivial when $p = 1$, we focus only on the cases of $p \ge 2$ in the following.
\medskip

For $p \ge 2$, \cite{DM2014} recently proposed a combinatorial algorithm, whose implementation runs faster and requires much less memory than that constructed on the breadth-first search algorithm. It turns out that their procedure has complexity $O(n^p)$. When $p = 2$, 3, the complexity of his procedure is of higher order than that of few existing algorithms; see for example \cite{RR1996a} and \cite{RS1998}.
\medskip

If carefully investigating Mozharovskyi's algorithm, it is easy to find that this proposal computes $D_{n}(0)$ actually relying on \eqref{DNPp1}. On the other hand, Proposition 1 indicates that we may utilize the fact that $D_{n}(0) = \min_{\mathbf{I}_{r} \in \mathcal{I}_{r}} D_{n}(\mathbf{S}_{\mathbf{I}_{r}}) - r/n$ to compute $D_{n}(0)$ with other $r\ (\neq  p - 1)$.
\medskip

Among $r = $1, 2, $\cdots$, $p - 2$, our favourite is
\begin{eqnarray}
\label{HDSp22}
    D_{n}(0) = \min_{\mathbf{I}_{p-2} \in \mathcal{I}_{p-2}} D_{n}(\mathbf{S}_{\mathbf{I}_{p-2}}) - \frac{p - 2}{n}.
\end{eqnarray}
The reasons are as follows. There are only ${n \choose p - 2}$ combinations $\mathbf{I}_{p-2}$. For each $\mathbf{I}_{p-2}$, $D_{n}(\mathbf{S}_{\mathbf{I}_{p-2}}) = D_{n, \mathbf{I}_{p-2}}(0)$, which is in fact a bivariate Tukey depth. While for bivariate data, it is known that some well-developed algorithms have only complexity $O(n \log (n))$ \citep{RR1996a}. In this sense, Mozharovskyi's algorithm can be further improved to the version of complexity $O(n^{p-1} \log(n))$. This motivates us to consider the following procedure.

\begin{algorithm} (for $p$-dimensional data with $p \ge 3$)\quad
\begin{enumerate}
    \item[] {\bf Input:} $\mathbf{X}^n = \{\mathbf{x}_1, \cdots ,\mathbf{x}_n\} \subset \mathcal{R}^p$, $3 \leq p < n < \infty$,  $\mathbf{X}^n \cup \{0\}$ in general position.

    \item[] \textbf{Step 1.} Let $N_{min} = n$. For each $\mathbf{I}_{p-2} := \{i_1, \cdots ,i_{p-2}\} \in \mathcal{I}_{p-2}$ (see \eqref{indxset}), do:
        \begin{enumerate}
            \item compute two orthogonal vectors $e_1$ and $e_2$ of the orthogonal complement subspace of that spanned by $\{\mathbf{x}_{i_1},\, \mathbf{x}_{i_2},\, \cdots, \, \mathbf{x}_{i_{p-2}}\}$,

            \item compute the bivariate Tukey depth $D_{n}(0)$ with respect to $\{\mathbf{y}_i := (e_1^{\top} \mathbf{x}_i,\, e_2^{\top} \mathbf{x}_i)^{\top}$, $i = 1,\, 2,\, \cdots,\, n\}$,

            \item if $N_{min} > n D_{n}(0)$, then $N_{min} = n D_{n}(0)$.
        \end{enumerate}

    \item[] \textbf{Step 2.} Return $\frac{N_{min} - (p - 2)}{n}$.
    \item[] {\bf Output:} $\frac{N_{min} - (p - 2)}{n}$.
\end{enumerate}
\end{algorithm}

Note that computing the bivariate Tukey depth is a quite key step in Algorithm 1, because it has to be repeatedly taken for ${n \choose p-2}$ times, which would be huge when $p$ and/or $n$ are large. Even a little improvement on the efficiency of the bivariate procedure may lead to a lot of CPU time saving.
To this end, we propose to consider the following approach, which can compute exactly the bivariate Tukey depth $D_{n}(0)$.

\begin{algorithm} (for bivariate data only)\quad
\begin{enumerate}
    \item[] {\bf Input:} $\mathbf{Y}^n = \{\mathbf{y}_1, \cdots ,\mathbf{y}_n\} \subset \mathcal{R}^2$.

    \item[] \textbf{Step 1.} Let $M_{min} = n$. Do:
        \begin{enumerate}
            \item compute $\{\widetilde{\mathbf{y}}_i\}_{i = 1}^m := \{\mathbf{y}\ |\ \mathbf{y} \neq 0,\, \mathbf{y} \in \mathbf{Y}^n\}$, where $m$ is the cardinality number,

            \item compute $\theta_i = -\widetilde{\mathbf{y}}_{i1} / \widetilde{\mathbf{y}}_{i2}$ for $\widetilde{\mathbf{y}}_i = (\widetilde{\mathbf{y}}_{i1},\, \widetilde{\mathbf{y}}_{i2})^\top$, $i = 1,\, 2,\, \cdots,\, m$,

            \item for $u_0 = (0,\, -1)^{\top}$, compute $(L_1, \, L_2,\, \cdots, L_m)$ and $S_0 = \min\{S_1$, $S_2\}$, where $S_1 = \sum_{i=1}^m L_i$, and $S_2 = m - S_1$ with $L_i = 1$ if $u_0^{\top} \widetilde{\mathbf{y}}_i \ge 0$ (actually, $\widetilde{\mathbf{y}}_{i2} \leq 0$), else $L_i = 0$ for $i = 1,\, \cdots,\, m$,

            \item if $M_{min} > S_0$, set $M_{min} = S_0$,

            \item compute the permutation $(j_1,\, j_2,\, \cdots,\, j_m)$ such that $\theta_{j_1}\leq \theta_{j_2} \leq \cdots \leq \theta_{j_m}$, and for each $k = 1:m$, do:

                \begin{enumerate}
                    \item[(i)] if $L_{j_k} = 0$, set $S_1 = S_1 + 1$, else $S_1 = S_1 - 1$,

                    \item[(ii)] compute $S_2 = m - S_1$, and update $S_0 = \min\{S_1,\, S_2\}$,

                    \item[(iii)] if $M_{min} > S_0$, let $M_{min} = S_0$.
                \end{enumerate}
        \end{enumerate}

    \item[] \textbf{Step 2.} Return $\frac{M_{min} + n - m}{n}$.
    \item[] {\bf Output:} $D_{n}(0) = \frac{M_{min} + n - m}{n}$.
\end{enumerate}
\end{algorithm}

In the literature, it is known that the sorting step is most time-consuming in computing the bivariate Tukey depth. Compared to the classical algorithm of \cite{RR1996a}, hereafter RR96, the efficiency of Algorithm 2 comes from two folds: (i) Algorithm 2 only needs to sort a sequence of length $m\ (m \leq n$, see Step 1-(e)), while that in RR96 is of length $2m$. Hence, the complexity $O(m\log(m))$ of Algorithm 2 is slightly better than that $O(2m\log(2m))$ of RR96. (ii) Algorithm 2 sorts directly the sequence $\{\theta_i\}_{i=1}^m$, rather than $\{\alpha_1,\, \alpha_2,\, \cdots,\, \alpha_m,\, \beta_1,\, \beta_2,\, \cdots,\, \beta_m\}$ as used by \cite[see pp. 519]{RR1996a}, where $\alpha_i \in [0, 2\pi)$ satisfy that $\tan(\alpha_i) = \theta_i$, and $\beta_i = \alpha_i + \pi$ if $\alpha_i \in [0, \pi)$, else $\beta_i = \alpha_i - \pi$ for $i = 1,\, \cdots,\, m$. Clearly, computing $\theta_i$'s is much simpler. For these reasons, we recommend to use it in Algorithm 1.

\vskip 0.1 in
\section{An adaptive iterative algorithm}
\paragraph{}
\vskip 0.1 in
\label{ADIA}

Most existing procedures have to fully address all ${n \choose p - 1}$ critical direction vectors, no matter where the point $z$ is located at. On the other hand, a great proportion of these vectors may be redundant, because when computing the Tukey depth, we are computing for the infimum of $P_n (u^{\top} X \leq u^{\top} z)$ with respect to $u$.
\medskip

\begin{figure}[H]
\captionsetup{width=0.85\textwidth}
\centering
\includegraphics[angle=0,width=3.5in]{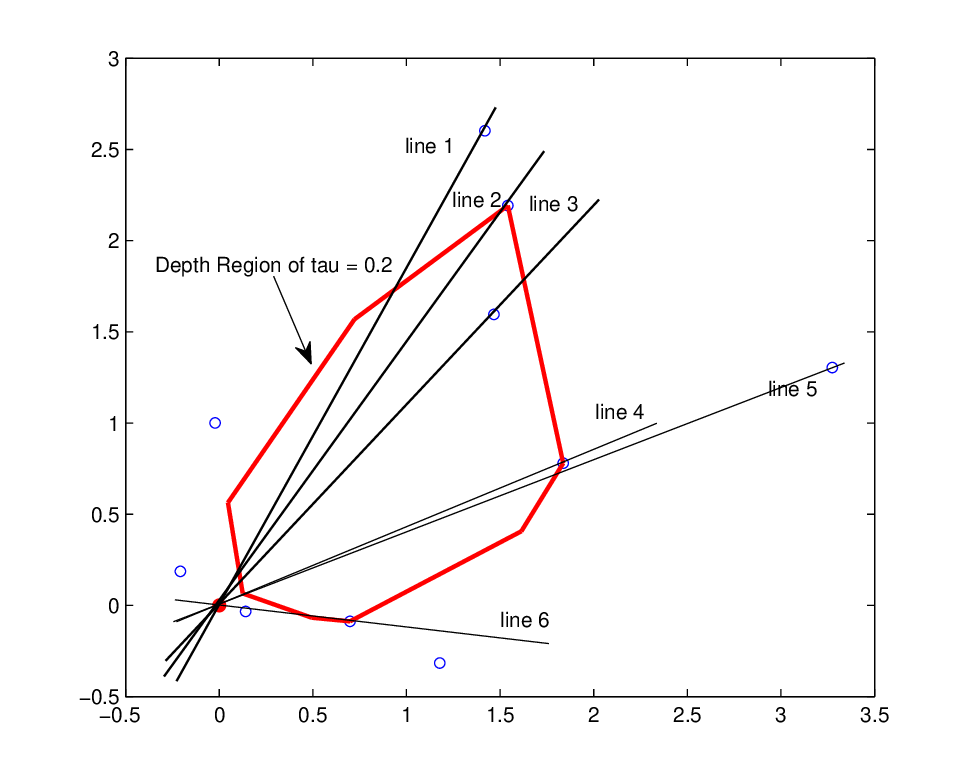}
\caption{Shown is an illustration of redundant direction vectors for computing the Tukey depth. Clearly, the critical direction vectors normal to Lines 1-6 are redundant if we have known that the depth of 0 is at most $0.2$.}
\label{fig:RDV}
\end{figure}

This may easily be seen from Figure \ref{fig:RDV}. In this illustration, we are computing the Tukey depth of 0 with respect to a data set containing 10 observations. Assume that we have known that an upper bound of the Tukey depth of 0 is $0.2$ through an approximate method. Then it is easy to conclude that critical direction vectors normal to Lines 1-6 are redundant, because using them can not produce a smaller depth value than 0.2. Hence it's better to eliminate them from consideration as many as possible. This idea seems to have been utilized by \cite{JKN1998} for \emph{bivariate} data.

In this section, we are interested to present an iterative procedure for dimensions $p \ge 3$. The most outstanding of this procedure is its ability to adaptively avoid considering many redundant critical direction vectors conditionally on the former iteration. Before proceeding further, let's provide two propositions as follows.
\medskip

\textbf{Proposition 2}. \textit{For any subspace $\mathbf{V}_r$ ($0 < r < p$) of $\mathcal{R}^p$ ($p > 1$), we have that}
\begin{eqnarray*}
    \sup_{x \in \mathbf{V}_r} D_{n}(x) \leq D_{n}(\mathbf{V}_r).
\end{eqnarray*}

\textbf{Proof}. This proposition can be proved as follows: Since the image of $P_n$ only can take a finite set of values: $0,\, 1/n,\, 2/n,\, \cdots,\, 1$, there must exist $u_0\in \mathcal{S}^{p-1} \cap \mathbf{V}_r^\bot$ such that, for any $x \in \mathbf{V}_r$,
\begin{eqnarray*}
    D_n(\mathbf{V}_r) = P_n(u_0^\top X \leq 0) = P_n(u_0^\top X \leq u_0^\top x) \ge D_n(x).
\end{eqnarray*}
This completes the proof.
\medskip

This proposition indicates that once $\mathbf{V}_r$ contains a point $x$ with $D_{n}(x) \ge \tau$, we must have $D_{n}(\mathbf{V}_r) \ge \tau$. In other words, if we known in advance that $D_n(0) < \tau$, then any subspace $\mathbf{V}$ such that $\mathbf{V} \cap \mathcal{D}_n(\tau) \neq \emptyset$ is redundant for computing $D_n(0)$, and may be eliminated, if possible, from consideration by the convexity of $\mathcal{D}_n(\tau)$, where $\mathcal{D}_n(\tau) = \{x\in \mathcal{R}^p\ |\ D_{n}(x) \ge \tau\}$ denotes the $\tau$-th Tukey depth region; see Figure \ref{fig:RDV} for an illustration.
\medskip

\textbf{Proposition 3}. \textit{Assume that $\mathbf{X}^n \cup \{0\}$ are in general position. For $p \ge 3$ and an any given combination $(\mathbf{x}_{i_1},\, \mathbf{x}_{i_2},\, \cdots,\, \mathbf{x}_{i_{p-2}})$, there are another observation $\mathbf{x}_{i_{p-1}}$ and $\bar{u}_0$ normal to the hyperplane passing through $\{0,\, \mathbf{x}_{i_1},\, \mathbf{x}_{i_2},\, \cdots,\, \mathbf{x}_{i_{p-2}},\, \mathbf{x}_{i_{p-1}}\}$ such that
\begin{eqnarray*}
    D_{n}(\mathbf{S}_{i_1,i_2,\cdots, i_{p-2}}) = P_n(\bar{u}_0^{\top} X \leq 0) - \frac{1}{n}.
\end{eqnarray*}
More importantly, for any $j_1,\, j_2,\, \cdots, j_{p-3}$ distinct and $j_1,\, j_2,\, \cdots, j_{p-3} \in \{i_1,\, i_2,\, \cdots,\, i_{p-2}\}$, it holds}
\begin{eqnarray*}
    D_{n}(\mathbf{S}_{i_1,i_2,\cdots, i_{p-2}}) \ge D_{n}(\mathbf{S}_{j_1,j_2,\cdots, j_{p-3}, i_{p-1}}).
\end{eqnarray*}

\textbf{Proof}. The first part can be proved trivially by following a similar fashion to that of Propositions 1-2. For the second part, since $\bar{u}_0 \in \mathbf{S}_{i_1,i_2,\cdots, i_{p-2}, i_{p-1}}^{\bot}$, then $\bar{u}_0 \in \mathbf{S}_{j_1,j_2,\cdots, j_{p-3}, i_{p-1}}^{\bot}$ holds for any $j_1,\, j_2,\, \cdots, j_{p-3} \in \{i_1,\, i_2,\, \cdots,\, i_{p-2}\}$. Using this, we obtain
\begin{eqnarray*}
    D_{n}(\mathbf{S}_{i_1,i_2,\cdots, i_{p-2}}) &=& P_n(\bar{u}_0^{\top} X \leq 0) - \frac{1}{n}\\
     &\ge& \min_{l \in \{1,\, 2,\, \cdots,\, n\} \setminus \{j_1,j_2,\cdots, j_{p-3}, i_{p-1}\}} P_n(\bar{u}_l^{\top} X \leq 0) - \frac{1}{n} \\
     &=& D_{n}(\mathbf{S}_{j_1,j_2,\cdots, j_{p-3}, i_{p-1}}),
\end{eqnarray*}
where $\bar{u}_l$ is the direction vector determined by $\{0,\, \mathbf{x}_{j_1},\, \mathbf{x}_{j_2},\, \cdots, \mathbf{x}_{j_{p-3}},\, \mathbf{x}_{i_{p-1}}\} \cup \{\mathbf{x}_l\}$.
\medskip

Proposition 3 is in fact telling us a way how to adaptively find the next subspaces possessing a smaller Tukey depth conditionally on the current $\mathbf{S}_{i_1,i_2,\cdots, i_{p-2}}$. It, together with Proposition 2 and \eqref{HDSp22},
motivates us to consider the following iterative procedure. Here we assume $n > 2p$. For $n \leq 2p$, we recommend to utilize directly Algorithm 1 to compute the depth value.


\begin{algorithm} \quad
\begin{enumerate}
    \item[] {\bf Input:} $\mathbf{X}^n = \{\mathbf{x}_1, \cdots ,\mathbf{x}_n\} \subset \mathcal{R}^p$, $3 \leq p$, $2p < n < \infty$,  $\mathbf{X}^n \cup \{0\}$ in general position.

    \item[] \textbf{Step 1.} Set $d_0 = 1$, $\mathcal{Q} = \emptyset$, $\mathcal{N} = \emptyset$.

    \item[] \textbf{Step 2.} Compute $u_j = \mathbf{x}_j / \|\mathbf{x}_j\|$, $j = 1,\, \cdots,\, n$. Set $\mathbf{U}_0 = \{u_1,\, -u_1,\, \cdots,\, u_n, -u_n\}$, do:

        \begin{enumerate}
        \item find $u_{\min} \in \mathbf{U}_0$ such that $u_{\min} = \arg \min_{u \in \mathbf{U}_0} P_n (u^{\top}X \leq 0)$,
        \item compute the permutation $(i_{1,0},\, i_{2,0},\, \cdots,\, i_{k,0},\, i_{k+1,0},\, \cdots,\,i_{n,0})$ such that
        \begin{eqnarray*}
            u_{\min}^{\top} \mathbf{x}_{i_{1,0}} \leq u_{\min}^{\top} \mathbf{x}_{i_{2,0}} \leq \cdots \leq u_{\min}^{\top} \mathbf{x}_{i_{k,0}} < 0 \leq u_{\min}^{\top} \mathbf{x}_{i_{k + 1,0}} \leq \cdots \leq u_{\min}^{\top} \mathbf{x}_{i_{n,0}},
        \end{eqnarray*}
        if $k$ does not exist, set $d_0 = 0$ and goto Step 6,

        \item find $\mathbf{x}_{i_{p-1,0}}$ such that $D_{n}(\mathbf{S}_{i_{k+1,0}, i_{k+2,0},\cdots, i_{k+p-2,0}}) = P_n(u_0^{\top} X \leq 0) - 1/n$ based on Algorithm 2, where $u_0$ is determined by $\{0,\, \mathbf{x}_{i_{k+1,0}}$, $\mathbf{x}_{i_{k+2,0}},\cdots,\, \mathbf{x}_{i_{k+p-2,0}},\, \mathbf{x}_{i_{p-1,0}}\}$, set $\mathbb{C}.index = \{i_{k + 1,0}, \cdots, i_{k + p - 2,0},\, i_{p-1,0}\}$ and $\mathbb{C}.depth = D_{n}(\mathbf{S}_{i_{k+1,0}, i_{k+2,0},\cdots, i_{k+p-2,0}}) - (p - 2) / n$. Here $\mathbb{C}$ is of the type \emph{struct} having two fields, namely, $index$ and $depth$.
        \end{enumerate}

    \item[] \textbf{Step 3.} (a) Push $\mathbb{C}$ into both $\mathcal{Q}$ and $\mathcal{N}$, (b) if $d_0 > \mathbb{C}.depth$, set $d_0 = \mathbb{C}.depth$.

    \item[] \textbf{Step 4.} Pop a $\mathbb{Q}$ from $\mathcal{N}$, and

        \begin{enumerate}
        \item for each $\{j_1,\, j_2,\, \cdots,\, j_{p-2}\} \subset \mathbb{Q}.index$, do:
            \begin{enumerate}
                \item[(i)] compute $d_{temp} = D_{n}(\mathbf{S}_{j_1, j_2, \cdots, j_{p-2}}) - (p - 2) / n$ by Algorithm 2,

                \item[(ii)] store in $\mathcal{T}$ all $h \in \{1,\, 2,\, \cdots,\, n\} \setminus \mathbb{Q}.index$ such that $\{0,\, \mathbf{x}_{j_1},\, \mathbf{x}_{j_2},$ $\cdots,\, \mathbf{x}_{j_{p-2}}$, $ \mathbf{x}_{h}\}$ determine a $u_0$ satisfying $P_n (u_0^{\top}X \leq 0) - 1/n = D_{n}(\mathbf{S}_{j_1, j_2, \cdots, j_{p-2}})$,

                \item[(iii)] for each $t \in \mathcal{T}$, do:

                \begin{enumerate}
                    \item[(A)] set $\mathbb{N}.index = \{j_1,\, j_2,\, \cdots,\, j_{p-2},\, t\}$ and $\mathbb{N}.depth = d_{temp}$,

                    \item[(B)] if $\mathbb{N} \notin \mathcal{Q}$, push $\mathbb{N}$ into both $\mathcal{Q}$ and $\mathcal{N}$,
                \end{enumerate}

                \item[(iv)] if $d_0 > d_{temp}$, set $d_0 = d_{temp}$, break Step 4(a) and goto Step 4(b),
            \end{enumerate}

                \item delete all $\mathbb{D}$ in both $\mathcal{Q}$ and $\mathcal{N}$ such that $\mathbb{D}.detph > d_0$,

                \item if $\mathcal{N} \neq \emptyset$, iterate Step 4, else goto Step 5.
        \end{enumerate}

    \item[] \textbf{Step 5.} (a) Compute $\{\mathbf{x}_{k_l}\}_{l=1}^{s} := \bigcup_{\mathbb{F} \in \mathcal{Q}} \mathbf{X}_{\mathbb{F}}$, where $\mathbf{X}_{\mathbb{F}} = \{\mathbf{x} \in \mathbf{X}^n| u_{\mathbb{F}}^\top \mathbf{x} \leq u_{\mathbb{F}}^\top X_{i_1^*}\}$ for $\mathbb{F}.index = \{i_1^*,\, i_2^*,\, \cdots,\, i_{p-1}^*\}$, $u_{\mathbb{F}}$ is determined by $\{0,\, X_{i_1^*},\cdots,\, X_{i_{p-1}^*}\}$ and satisfies that $P_n(u_{\mathbb{F}}^\top X \leq u_{\mathbb{F}}^\top X_{i_1^*}) = \min\{P_n(u_{\mathbb{F}}^\top X \leq u_{\mathbb{F}}^\top X_{i_1^*}),\, P_n(-u_{\mathbb{F}}^\top X \leq -u_{\mathbb{F}}^\top X_{i_1^*})\}$.
    (b) Likewise to Algorithm 1, for each $\{i_1, \cdots ,i_{p-2}\} \in \widetilde{\mathcal{I}}_{p-2} =\{\{j_1, j_2,\cdots, j_{p-2}\}\ |\ j_1, j_2,$ $\cdots, j_{p-2}$ distinct, and $j_1, j_2,\cdots, j_{p-2} \in \{k_1, k_2,\cdots, k_s\}\}$, do:
        \begin{enumerate}
            \item[(i)] compute $d_{temp} = D_{n}(\mathbf{S}_{i_1, i_2, \cdots, i_{p-2}}) - (p - 2) / n$,

            \item[(ii)] if $d_0 > d_{temp}$, set $d_0 = d_{temp}$.
        \end{enumerate}

    \item[] \textbf{Step 6.} Return $d_0$.
    \item[] {\bf Output:} $D_{n}(0) = d_0$.
\end{enumerate}
\end{algorithm}

In Algorithm 3, Step 2 serves mainly for computing an upper bound $d_0$ for the Tukey depth and an initial $\mathbf{S}_{i_{k+1,0}, i_{k+2,0},\cdots, i_{k+p-2,0}}$; see also \cite{RS1998, CN2008} for some other approximate procedures, which may be used as an alterative here. 
The direction vectors considered in Step 2 are useful in reducing the computational burden when $D_{n}(0)$ is small. Steps 4-5 are key steps of Algorithm 3. Since Proposition 3 guarantees that the Tukey depth of each $\mathbf{S}_{j_1, j_2, \cdots, j_{p-2}}$ considered in Step 4(a) is no larger than that of $\mathbf{S}_{i_1,i_2,\cdots, i_{p-2}}$, a great proportion of critical direction vectors would be adaptively eliminated from the computation.
\medskip

In Step 2, we only use $n$ fixed direction vectors. Hence, the complexity of this step is $O(n\log(n))$. In fact, provided that no more than than $n^{p-1}$ direction vectors are utilized, the complexity would be $\leq O(n^{p - 1} \log (n))$. Next, according to the principle of this algorithm, Step 4 traverses the possible combinations $\{j_1,\, j_2,\, \cdots,\, j_{p-2}\}$ \emph{without repetition}. Since not all such combinations would be traversed, the complexity of Step 4 is $\leq O(n^{p - 1} \log (n))$. A similar situation applies to Step 5. Hence, the whole complexity of Algorithm 3 is $\leq  O(n^{p - 1} \log (n))$. Furthermore, based on the former step, we update timely in Step 4(b) both $\mathcal{Q}$ and $\mathcal{N}$ by deleting many
entities $\mathbb{D}$. Hence, Algorithm 3 requires quite minimal memory. 
\medskip


\vskip 0.1 in
\section{Performances}
\paragraph{}
\vskip 0.1 in
\label{Comparisons}

In this section, we will conduct a few data examples to investigate the performance of the proposed algorithms. All of these results are obtained on a HP Pavilion dv7 Notebook PC with Intel(R) Core(TM) i7-2670QM CPU @ 2.20GHz, RAM 6.00GB, Windows 7 Home Premium and Matlab 7.8.

\subsection{Illustrations}
\paragraph{}

In this subsection, we are interested to illustrate the performance of the proposed algorithms in terms of both computation time and accuracy based on the real data. For the sake of comparison, we also report the results obtained by the combinatorial algorithm developed by \cite{DM2014}, and the naive algorithm proposed by \cite{LZ2014a}. For convenience, in the sequel we denote the refine combinatorial algorithm as \emph{RCom}, the adaptive iterative algorithm as \emph{ADIA}, and the algorithms of \cite{DM2014} and \cite{LZ2014a} as \emph{DM14} and \emph{LZ14}, respectively.
\medskip

Two data sets are considered in the following. The first data set is taken from \cite{HS2007}, and has been investigated by \cite{LZ2014a} as an illustration. It consists of 64 samples as a part of a evolution of the vocabulary of children obtained from a cohort of pupils from the eighth through 11th grade levels. The second data set is a part of the the daily simple returns of IBM stock from 1970 January 01 to 2008 December 25 used by \cite{Tsa2010}. It currently can be downloaded from his teaching page:  \url{http://faculty.chicagobooth.edu/ruey.tsay/teaching/fts3/d-ibm3dx7008.txt}. The original data set consists of 755 observations. \emph{Remarkably}, our goal here is not to perform a thorough analysis for data, but rather to show how the algorithms work in practice.

\begin{table}[H]
{\scriptsize
\begin{center}
    \captionsetup{width=0.85\textwidth}
    \caption{Computation time (in seconds).}
    \label{Tab:NumTime}
    \begin{tabular}{p{0.1cm}p{0.4cm}p{1.0cm}rrrrrr}
    \toprule
    $p$ & Data & $n$ &\multicolumn{6}{l}{Computation time} \\
    \cmidrule(r){4-9}
                         &&&\multicolumn{1}{c}{\emph{RCom}} &\multicolumn{1}{c}{\emph{ADIA$_{\min}$}}&\multicolumn{1}{c}{\emph{ADIA}$_{\text{mean}}$}&\multicolumn{1}{c}{\emph{ADIA}$_{\max}$} &\multicolumn{1}{c}{\emph{DM14}} &\multicolumn{1}{c}{\emph{LZ14}}  \\
    \midrule
    3  & Voc&  64 &   0.0145 & 0.0066  &  0.0117 &  0.0380 &   0.1108  &    2.7740 \\[0.6ex]
       & IBM&  200&   0.0908 & 0.0146  &  0.0387 &  0.1287 &   1.1892  &   31.9022 \\[0.6ex]
       & IBM&  500&   0.4925 & 0.0283  &  0.1472 &  0.5609 &   8.2659  &  285.4298 \\[1.6ex]
    4  & Voc&  64 &   0.4614 & 0.0335  &  0.1410 &  0.3731 &   2.9491  &   89.8435 \\[0.6ex]
       & IBM&  200&   8.8961 & 0.0620  &  0.4624 &  3.3255 &  86.4703  & 3476.2415 \\[0.6ex]
       & IBM&  500& 120.0689 & 0.1411  &  2.4635 & 21.9474 & 1721.5922 & $--\quad$\\[0.6ex]
    \bottomrule
    \end{tabular}
\end{center}}
\end{table}

\begin{figure}[H]
\centering
	\subfigure[Depth (Voc, $p = 3$, $n = 64$)]{
	\includegraphics[angle=0,width=1.75in]{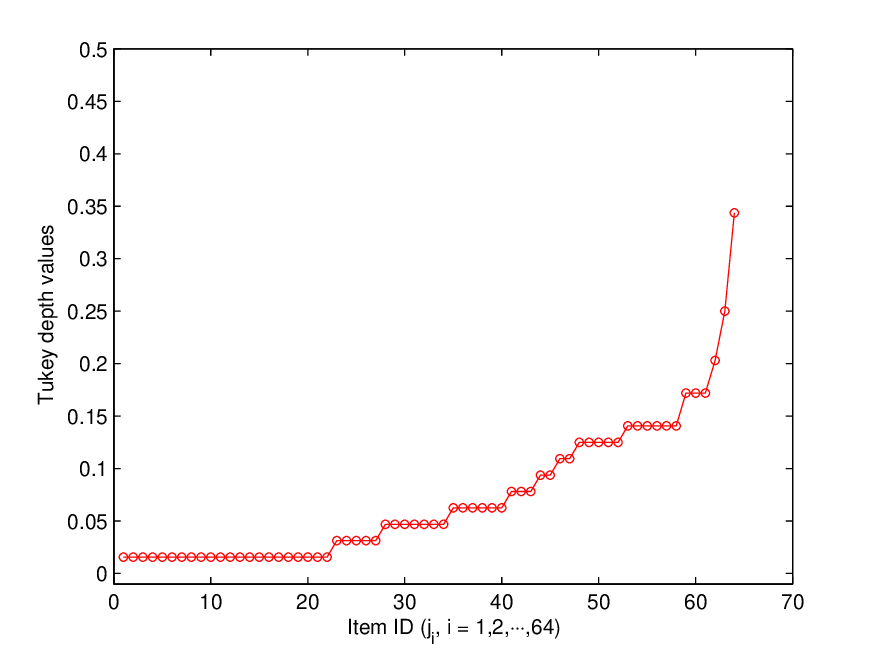}
	\label{HD2D600exp12}
	}\quad
	\subfigure[Depth (IBM, $p = 3$, $n = 200$)]{
	\includegraphics[angle=0,width=1.75in]{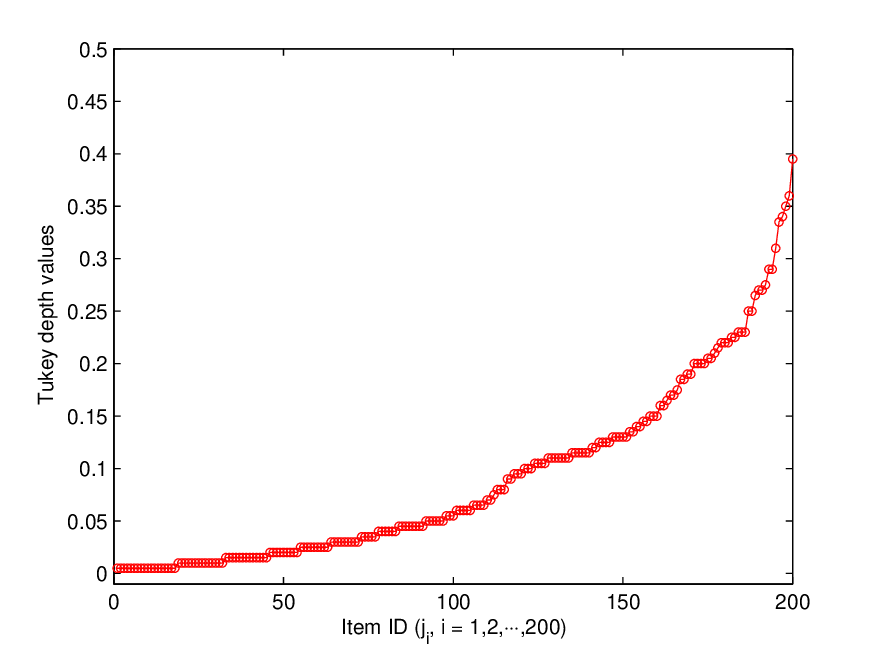}
	\label{HD2D600norm}
	}\quad
	\subfigure[Depth (IBM, $p = 3$, $n = 500$)]{
	\includegraphics[angle=0,width=1.75in]{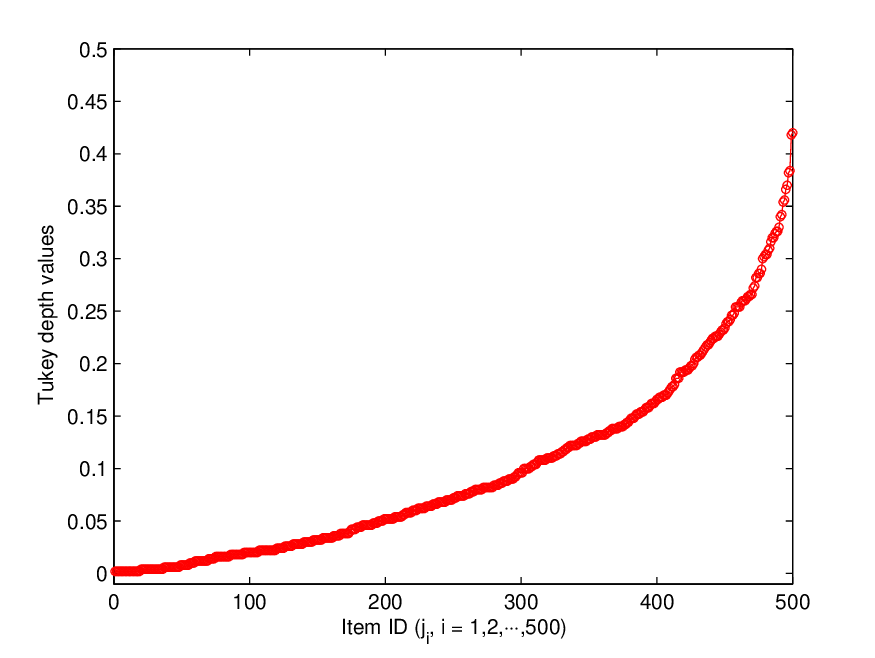}
	\label{PD2D600exp12}
	}\quad
	\subfigure[Time (Voc, $p = 3$, $n = 64$)]{
	\includegraphics[angle=0,width=1.75in]{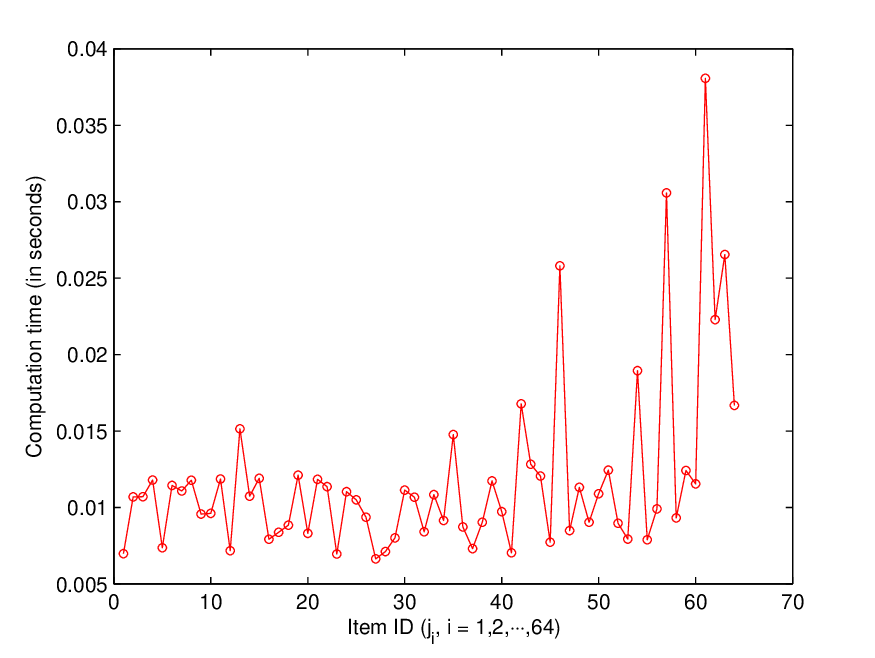}
	\label{PD2D600norm}
	}\quad
	\subfigure[Time (IBM, $p = 3$, $n = 200$)]{
	\includegraphics[angle=0,width=1.75in]{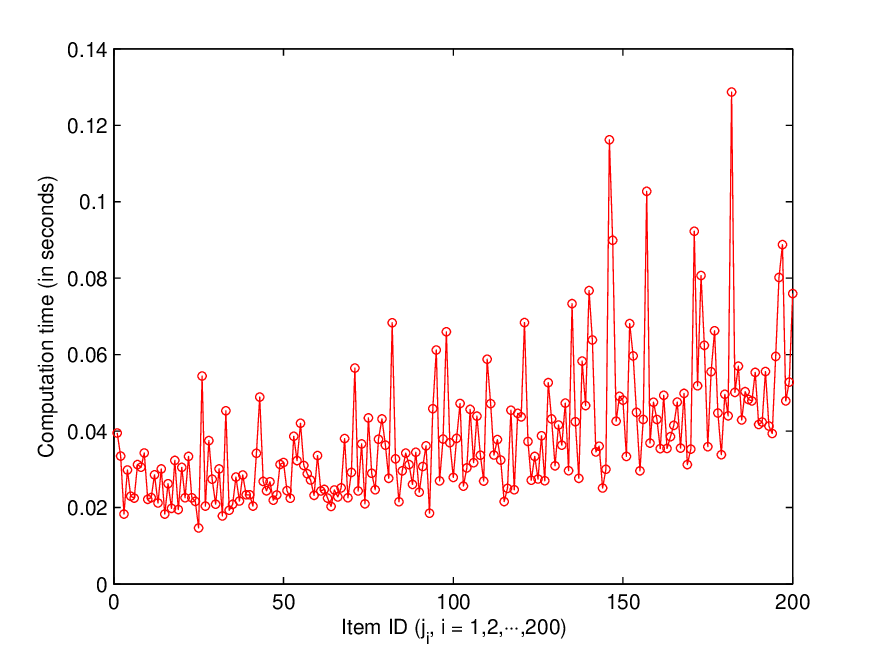}
	\label{MPD2D600exp12}
	}\quad
	\subfigure[Time (IBM, $p = 3$, $n = 500$)]{
	\includegraphics[angle=0,width=1.75in]{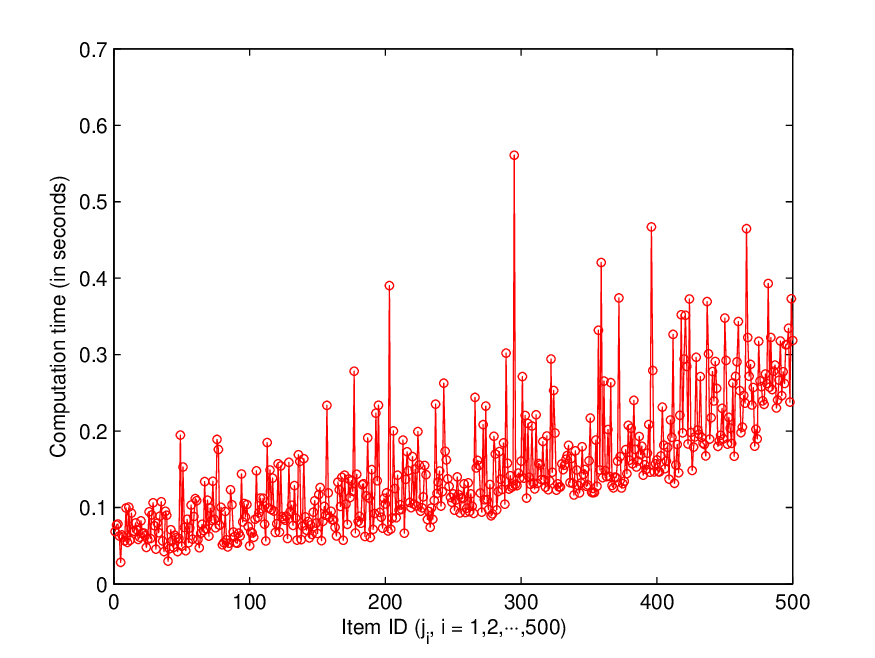}
	\label{MPD2D600norm}
	}\quad
	\subfigure[Depth (Voc, $p = 4$, $n = 64$)]{
	\includegraphics[angle=0,width=1.75in]{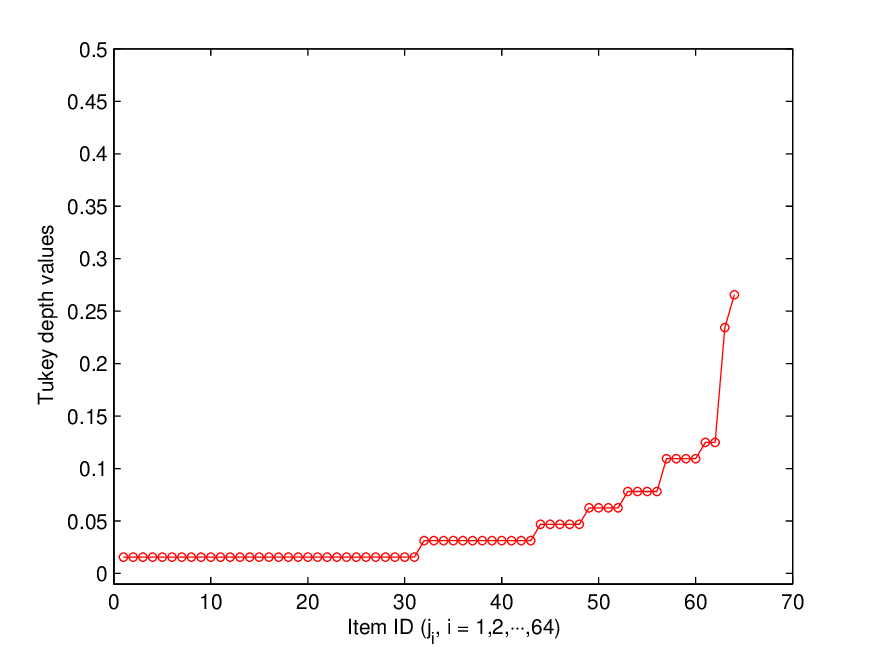}
	\label{HD2D600exp12}
	}\quad
	\subfigure[Depth (IBM, $p = 4$, $n = 200$)]{
	\includegraphics[angle=0,width=1.75in]{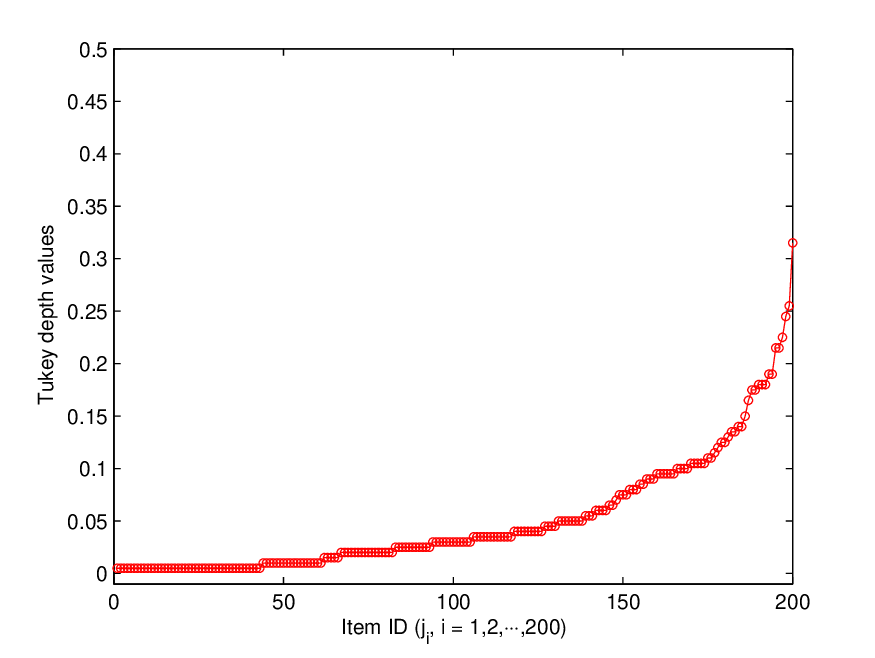}
	\label{HD2D600norm}
	}\quad
	\subfigure[Depth (IBM, $p = 4$, $n = 500$)]{
	\includegraphics[angle=0,width=1.75in]{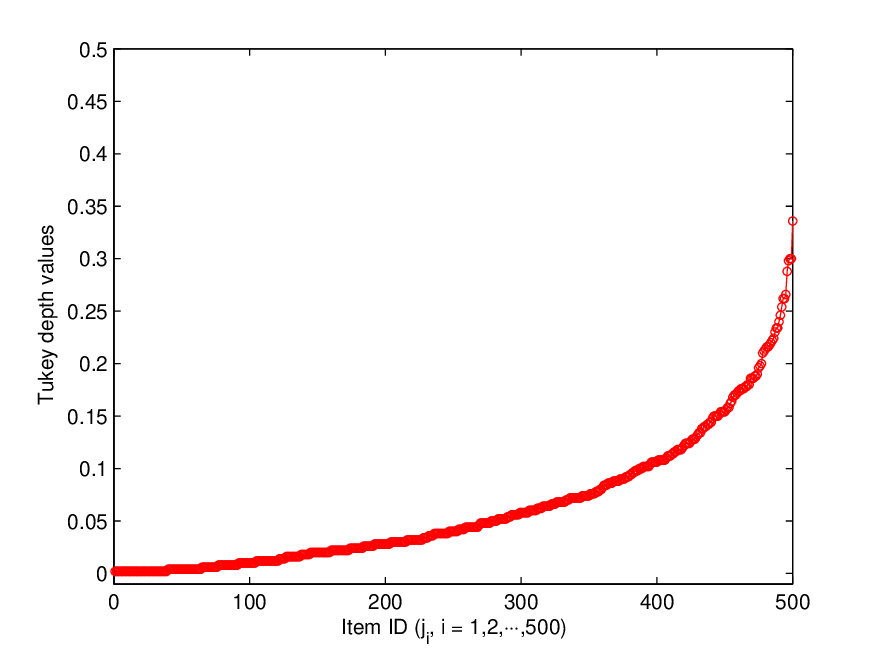}
	\label{PD2D600exp12}
	}\quad
	\subfigure[Time (Voc, $p = 4$, $n = 64$)]{
	\includegraphics[angle=0,width=1.75in]{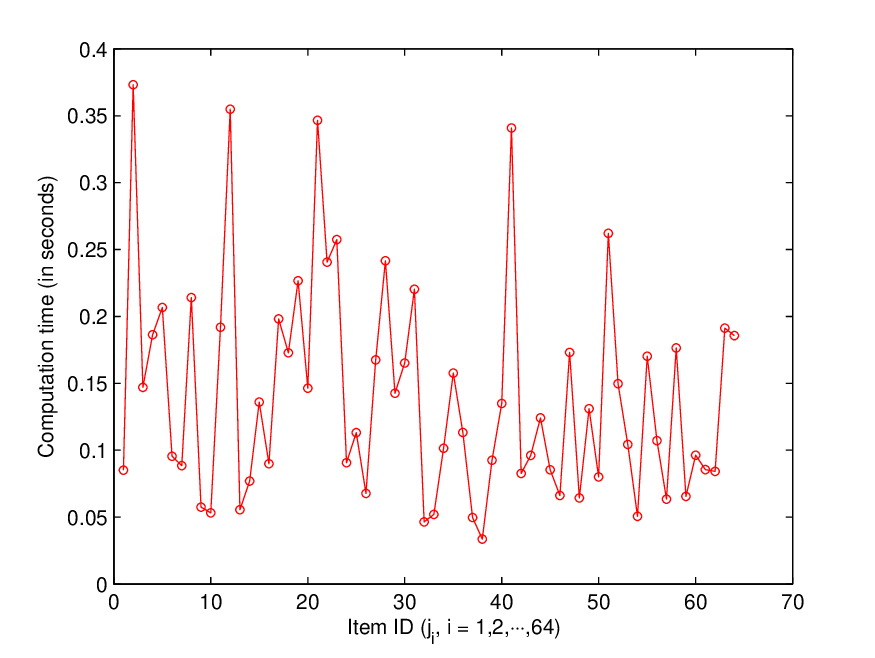}
	\label{PD2D600norm}
	}\quad
	\subfigure[Time (IBM, $p = 4$, $n = 200$)]{
	\includegraphics[angle=0,width=1.75in]{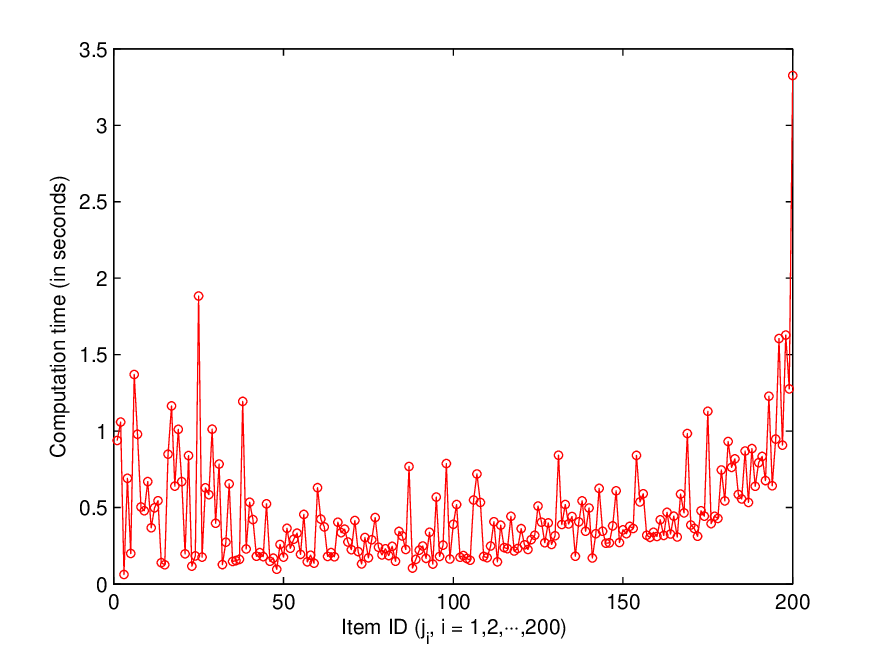}
	\label{MPD2D600exp12}
	}\quad
	\subfigure[Time (IBM, $p = 4$, $n = 500$)]{
	\includegraphics[angle=0,width=1.75in]{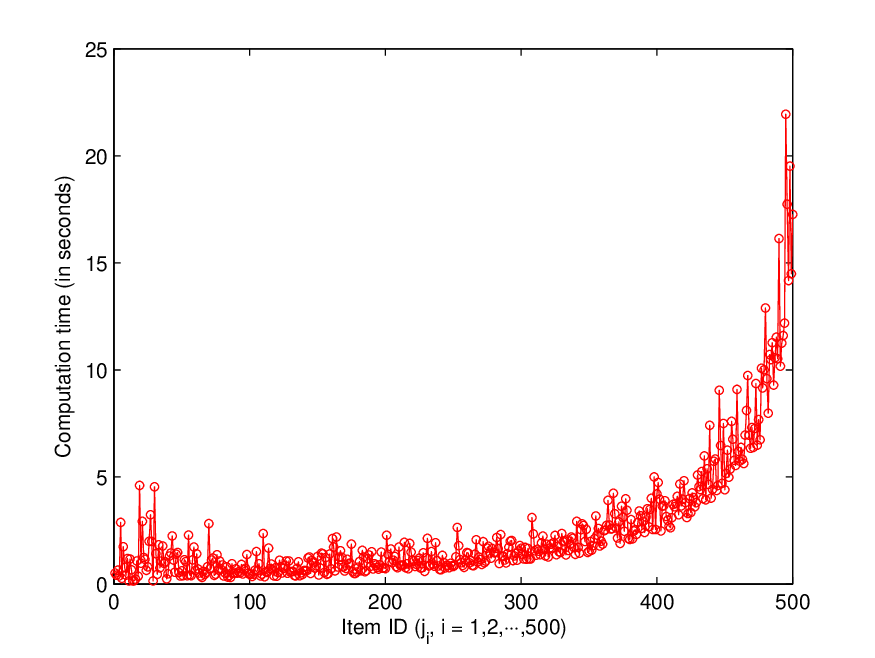}
	\label{MPD2D600norm}
	}
\caption{Shown are the Tukey depths (sorted ascending) and the corresponding computation times (in seconds) of the observations of Voc and IBM consumed by \emph{ADIA}.}
\label{fig:CompTime}
\end{figure}

Both data sets are 4-dimensional. For each observation, both proposed algorithms compute its
\emph{exact} Tukey depth, which coincide with those computed by \emph{DM14} and \emph{LZ14}. Table \ref{Tab:NumTime} reports the computation time (in seconds) for calculating the Tukey depth of a single observation. Here Voc denotes the vocabulary data, and IBM stands for the IBM stock data. $p = 3$ means we only use the first three columns of the data set, and $n = 200$ the first 200 rows. The sign `$--$' in Table \ref{Tab:NumTime} means this depth value is not computable in 8 hours. Since all \emph{Rcom}, \emph{DM14} and \emph{LZ14} have to fully address ${n \choose p - 1}$ critical direction vectors for every observation, the time for calculating each observation is almost the same. We only list the average computation time. Whereas the computation time consumed by \emph{ADIA} depends on the Tukey depth of the point being computing for a given data set, and therefore we report additionally its minimum, mean and maximum computation time (under the titles \emph{ADIA}$_{\min}$, \emph{ADIA}$_{\text{mean}}$ and \emph{ADIA}$_{\max}$, respectively). The smaller the depth of the observation being calculating is, the less the computation time \emph{ADIA} tends to consume; see Figure \ref{fig:CompTime} for more details.

Table \ref{Tab:NumTime} indicates that both the proposed algorithms run much faster than the existing algorithms. Among them, the implementation of \emph{ADIA} tends to run most the fastest when $n$ and/or $p$ are large. It requires no more than 3 seconds (in average) to obtain the depth of a single point in all illustrations here.
\medskip

It is worth mentioning that the algorithm of \cite{DM2014} is implemented here by us in Matlab for convenience of comparison. It appears to be slower than what was reported in \cite{DM2014}. This is possible, because their computations are based on C++, which usually runs faster than Matlab, especially when there are a great number of iterations involved.

\subsection{Speed comparisons}
\paragraph{}

In the following, we further compare the speeds of the proposed algorithms with that of \emph{DM14} based on the simulated data. The data are generated from the 3, 4, 5, 6-dimensional standard normal distributions with sample size $n =$ 40, 80, 160, $\cdots$, 2560. For each combination of $p \in \{3, 4, 5, 6\}$ and $n \in \{ 40, 80, 160, \cdots, 2560\}$, we compute repeatedly 10 times the Tukey depths of $z = \alpha \mathbf{1}_p$ with $\alpha = 0, 0.4, 0.8, 1.2$, where $\mathbf{1}_p$ denotes the $p$-dimensional vector of ones. We report the average computation time of \emph{RCom} and \emph{DM14} in Table \ref{AETD1}, and that of \emph{ADIA} in Table \ref{AETD2}, respectively. Since as pointed above, the computation time of both \emph{RCom} and \emph{DM14} do not depend on the Tukey depth of $z = 0$ being computing, we report only the average computation time corresponding to $\alpha = 0$ here.

\begin{table}[H]
\captionsetup{width=0.95\textwidth}
{\scriptsize
\begin{center}
    \caption{Average computation times (in seconds) of \emph{RCom} and \emph{DM14}.}
    \label{AETD1}
    \begin{tabular}{ccrrrrrrr}
    \toprule
     \multicolumn{1}{c}{$p$}   & \multicolumn{1}{c}{Method}    &\multicolumn{7}{c}{$n$} \\
\cmidrule(r){3-9}
    &               &\multicolumn{1}{c}{40}  &\multicolumn{1}{c}{80} &\multicolumn{1}{c}{160}   &\multicolumn{1}{c}{320} &\multicolumn{1}{c}{640} &\multicolumn{1}{c}{1280} &\multicolumn{1}{c}{2560}\\
    \midrule
     3    &\emph{RCom} & 0.0268  &   0.0481  &     0.0899 &   0.2531  &   0.8073  &   4.6438  &    11.7662   \\[0.6ex]
          &\emph{DM14} & 0.1466  &   0.8860  &     5.1369 &  35.0182  & 255.9098  &2072.8847  & 16230.1634   \\[1.8ex]
     4    &\emph{RCom} & 0.1876  &   0.8045  &     5.0550 &  43.3812  & 248.5797  &1878.8783  & 14681.9140   \\[0.6ex]
          &\emph{DM14} & 0.6377  &   5.7872  &    45.1497 & 393.0911  &3612.5004  & $--\quad$ &  $--\quad$    \\[1.8ex]
     5    &\emph{RCom} & 1.9517  &  21.1227  &   262.3251 &3864.4624  & $--\quad$ &$--\quad$  &  $--\quad$    \\[0.6ex]
          &\emph{DM14} & 5.8554  & 104.6715  &  1847.4277 & $--\quad$ & $--\quad$ &$--\quad$  &   $--\quad$   \\[1.8ex]
     6    &\emph{RCom} &18.4515  & 446.0153  & 14695.1294 & $--\quad$ & $--\quad$ &$--\quad$  &  $--\quad$    \\[0.6ex]
          &\emph{DM14} &48.6214  &1794.7799  &  $--\quad$ & $--\quad$ & $--\quad$ &$--\quad$  &   $--\quad$   \\[0.6ex]
    \bottomrule
    \end{tabular}
\end{center}}
\end{table}

Tables \ref{AETD1}-\ref{AETD2} indicate that both the proposed algorithms \emph{run much faster than} that of \emph{DM14}. By denoting $t(n, p)$ to be the computational time for the combination $(n, p)$, we can see that the value $\frac{t(2n, p)}{t(n, p)}$ corresponding to \emph{RCom} is $\approx 2^{p - 1} \log(2n) / \log(n)$, better than that of \emph{DM14} which is $\approx 2^p \log(2n) / \log(n)$. Nevertheless, for the combination of $(n, \alpha) = (5, 0.0)$ in Table \ref{AETD2}, the average time jumps from 404.51 to 13177.52 with $n$ increased from 320 to 640. (A similar observation could be seen with dimension 6 as $n$ moves from 80 to 160 at $\alpha = 0.0$.) Intuitively, it seems abnormal because $13177.52 / 404.51 \approx 32.58 \approx 2^{5} \log(640) / \log(320) > 2^{5-1} \log(640) / \log(320) \approx 17.92$. However, this does \textbf{\emph{not}} mean that the complexity of \emph{ADIA} would be $> O(n^{p-1}\log(n))$, although we are unable to obtain a precise order (even approximately) for the complexity of \emph{ADIA} at this moment. Our reason is that \emph{ADIA} probably saves more computational time relative to \emph{RCom} for the combination $(n, p) = (320, 5)$ than that for the combination $(n, p) = (640, 5)$ with $\alpha = 0.0$ in these 10 repeated computations. This results in $13177.52 / 404.51 \approx 32.58 \approx 2^{5} \log(640)> 2^{5-1} \log(640) / \log(320)$ though. The computational time of \emph{ADIA} is on the other hand much less than that of \emph{RCom}, whose empirical complexity is approximately $\approx 2^{p - 1} \log(2n) / \log(n)$, for each combination $(n, p)$ as indicated in Table \ref{AETD1}, nevertheless.

\begin{table}[H]
\captionsetup{width=0.95\textwidth}
{\scriptsize
\begin{center}
    \caption{Average computation times (in seconds) of \emph{ADIA}.}
    \label{AETD2}
    \begin{tabular}{cp{0.80cm}rrrrrrr}
    \toprule
     \multicolumn{1}{c}{$p$}   & \multicolumn{1}{l}{$\alpha$}    &\multicolumn{7}{c}{$n$} \\
\cmidrule(r){3-9}
    &               &\multicolumn{1}{c}{40}  &\multicolumn{1}{c}{80} &\multicolumn{1}{c}{160}   &\multicolumn{1}{c}{320} &\multicolumn{1}{c}{640} &\multicolumn{1}{c}{1280} &\multicolumn{1}{c}{2560}\\
    \midrule
   3&	0.0	& 0.0193   & 0.0634    &   0.0503    &   0.1297   &    0.4372   &    1.8128   &   9.1473\\[0.6ex]
   	&	0.4	& 0.0167   & 0.0470    &   0.0668    &   0.0968   &    0.3418   &    1.3816   &   5.3099\\[0.6ex]
   	&	0.8	& 0.0065   & 0.0131    &   0.0180    &   0.0460   &    0.3248   &    0.4342   &   5.4500\\[0.6ex]
   	&	1.2	& 0.0067   & 0.0136    &   0.0186    &   0.0390   &    0.1220   &    0.3514   &   1.3141\\[1.8ex]
   4&	0.0	& 0.3416   & 0.3211    &   1.7991    &   6.5374   &   60.6911   &  408.8870   &5570.4364\\[0.6ex]
   	&	0.4	& 0.0607   & 0.1704    &   0.5131    &   1.7255   &   11.9404   &  105.2564   & 740.5502\\[0.6ex]
   	&	0.8	& 0.1070   & 0.1172    &   0.1140    &   0.5171   &    1.7526   &   10.1355   &  72.9394\\[0.6ex]
   	&	1.2	& 0.1373   & 0.1465    &   0.0538    &   0.2278   &    0.9409   &    2.2443   &   5.0238\\[1.8ex]
   5&	0.0	& 0.9466   & 6.9212    &  68.0360    & 404.5147   &13177.5227   & $--\quad$   &$--\quad$\\[0.6ex]
   	&	0.4	& 0.4048   & 1.0052    &   8.3791    &  43.6717   &  609.6461   &11022.4774   &$--\quad$\\[0.6ex]
   	&	0.8	& 1.2751   & 1.1352    &   1.6600    &   5.3665   &   15.8776   &  105.1733   &1749.6048\\[0.6ex]
   	&	1.2	& 0.1923   & 0.1001    &   0.3339    &   0.3162   &    1.5744   &   47.5422   &  91.6245\\[1.8ex]
   6&	0.0	&12.3953   &95.3378    &3883.9385    &$--\quad$   & $--\quad$   & $--\quad$   &$--\quad$\\[0.6ex]
   	&	0.4	& 4.3607   &18.3686    & 173.8880    &1504.1087   & $--\quad$   & $--\quad$   &$--\quad$\\[0.6ex]
   	&	0.8	& 0.1638   & 0.2295    &  62.8009    & 158.7744   & 1777.3669   & $--\quad$   &$--\quad$\\[0.6ex]
   	&	1.2	& 1.1149   & 3.0121    &   1.4400    &   4.9680   &   19.3288   &   30.2152   &  75.0221\\[0.6ex]
    \bottomrule
    \end{tabular}
\end{center}}
\end{table}

\vskip 0.1 in
\section{Concluding discussions}
\paragraph{}
\label{Conclude}
\vskip 0.1 in

In this paper, we investigate the computing issue of the Tukey depth. To facilitate the discussions, we extend the conventional definition of the Tukey depth for a single point into the version for a subspace. Three propositions are provided. Proposition 1 finds a connection between $D_n(0)$ and a finite number of the Tukey depths of some $r$-dimensional subspaces spanned by observations, $r = 1,\, \cdots,\, p - 1$. Interesting in this proposition is the adjusted term $-r/n$, omitting which would lead to overestimation. A refined combinatorial algorithm, i.e., \emph{RCom}, is constructed on this proposition. It has complexity $O(n^{p-1}\log(n))$.
\medskip

Proposition 2 explains why we can eliminate some critical direction vectors from consideration, while Proposition 3 tells how to avoid considering them. These two propositions are simple, but useful in computing the Tukey depth. The reason is that the Tukey depth is defined to be the infimum of $P_n (u^{\top} X \leq u^{\top} z)$ with respect to $u$, and many critical direction vectors have no contribution to the final result. Based on these ideas, we propose the second algorithm, namely, \emph{ADIA}. Unlike \emph{Rcom}, \emph{ADIA} does not take accounts of all critical direction vectors. Hence, its complexity is $\leq O(n^{p-1}\log(n))$. It turns out that the computation time of \emph{ADIA} is depth-depending, and it runs very fast if the Tukey depth of $z$ is small. In all the experiments we conducted, using \emph{ADIA} obtains the exact depth values.
\medskip


As mentioned by \cite{DM2014}, there are many other depth notions being of both projection and quasiconcave properties, such as the projection depth \citep{Zuo2003} and the zonoid depth \citep{KM1997}. Efficient algorithms for these depths exist only for bivariate data; see, e.g., \cite{LZ2014b}. Therefore, how to utilize the quasiconcave of these depth notions as did in this paper to reduce the computational burden in higher dimensions is still worthy of further consideration.

\vskip 0.1 in
\section*{Acknowledgments}
\paragraph{}
\vskip 0.1 in


The author thanks Prof. Mosler, K. and Dr. Mozharovskyi, P. for their valuable discussions
during the preparation of this manuscript. The author also greatly appreciates two anonymous reviewers for their careful reading and insightful comments, which led to many improvements in this paper. This research is supported by NSFC of China (No. 11601197, 11461029, 71463020), the NSF of Jiangxi Province (No. 20161BAB201024, 20151BAB211016), and the Key Science Fund Project of Jiangxi provincial education department (No. GJJ150439).

\vskip 0.1 in
\section*{Compliance with Ethical Standards}
\paragraph{}
\vskip 0.1 in

I am the sole author of this manuscript. This research involves no human participants and/or animals, and has no conflict of interest.

\end{document}